\documentclass[prd,longbibliography,superscriptaddress,showpacs,showkeys,twocolumn,floatfix,eqsecnum]{revtex4-1}
\usepackage{textcomp}
\usepackage{eurosym}
\usepackage{amsfonts}
\usepackage{url}
\usepackage{array}
\usepackage{amsthm}
\usepackage{bm}
\usepackage{palatino}
\usepackage[colorinlistoftodos]{todonotes}
\usepackage{mathpazo}
\usepackage{supertabular}
\usepackage{subfig}
\usepackage{amssymb}
\usepackage{eurosym}
\usepackage{amsmath}
\usepackage{graphics}
\usepackage{color}
\usepackage{graphicx}
\usepackage[font={footnotesize,it}]{caption}
\usepackage[colorlinks=true,
            linkcolor=blue,
            urlcolor=blue,
            citecolor=green]{hyperref}
\usepackage{tabstackengine}
\usepackage{graphicx}
\usepackage[english]{babel}
\usepackage{amsfonts}
\usepackage{eurosym}
\usepackage{calrsfs}
\usepackage{color}

\def\be{\begin{equation}}
\def\ee{\end{equation}}

\begin{document}

\title{Determining the topology and deflection angle of ringholes via Gauss-Bonnet theorem }

\author{Kimet Jusufi}
\email{kimet.jusufi@unite.edu.mk}
\affiliation{Physics Department, State University of Tetovo, Ilinden Street nn, 1200,
Tetovo, North Macedonia.}

\begin{abstract}
In this letter, we use a recent wormhole metric known as a ringhole [Gonzalez-Diaz, Phys.\ Rev.\ D {\bf 54}, 6122, 1996] to determine the surface topology and the deflection angle of light in the weak limit approximation using the Gauss-Bonnet theorem (GBT). We apply the GBT and show that the surface topology at the wormhole throat is indeed a torus by computing the Euler characteristic number. As a special case of the ringhole solution, one can find the Ellis wormhole which has the surface topology of a 2-sphere at the wormhole throat.  The most interesting results of this paper concerns the problem of gravitational deflection of light in the spacetime of a ringhole geometry by applying the GBT to the optical ringhole geometry.  It is shown that, the deflection angle of light depends entirely on the geometric structure of the ringhole geometry encoded by the parameters $b_0$ and $a$, being the ringhole throat radius and the  radius of the circumference generated by the circular axis of the torus, respectively. As special cases of our general result, the deflection angle by Ellis wormhole is obtained. Finally, we work out the problem of deflection of relativistic massive particles and show that the deflection angle remains unaltered by the speed of the particles.
\end{abstract}

\date{\today}
\maketitle

\affiliation{Physics Department, State University of Tetovo, Ilinden Street nn, 1200,
Tetovo, Macedonia.} 
\affiliation{Institute of Physics, Faculty of Natural Sciences and Mathematics, Ss. Cyril
and Methodius University, Arhimedova 3, 1000 Skopje, Macedonia.}

{} 

\section{Introduction}
Solving the Einstein's field equations of general relativity leads to very interesting solutions which can be interpreted as tunnel-like structures connecting two different spacetime regions, known as wormholes. The concept of wormholes was introduced by Flamm in 1916 \cite{Flamm}, soon after the first black hole exact solution was found. After that, Einstein and Rosen \cite{Einstein} introduced a coordinate transformation which eliminates the curvature singularity, this lead to the famous bridge-like structure connecting two spacetime regions known as Einstein-Rosen bridges. In fact, they tried to interpret these solutions as a model for elementary particles, but that idea turned out to be unsuccessful. Afterwards, Wheeler carried out a pioneering work in wormhole physics \cite{fw,wh1,Wh2,Wh3}, followed by important contributions of Ellis, Bronnikov, Clement, Morris, and Thorne \cite{Ellis1,Ellis2,br1,clm,Morris}. The main difficulty underlying these objects relies on the fact that the wormhole geometric structure involves a strange type of matter known as the exotic matter. Exotic matter can be described by a stress-energy tensor components which violates the null, weak and strong energy conditions at the wormhole throat \cite{Visser}.  As of today, however, there is no experimental evidence supporting their existence in nature, i.e wormholes have never been observed. Despite this fact, wormholes have attracted broad interest in the scientific community and, among various aspects we point out the gravitational lensing effect \cite{Teo,lobo1,lobo2,branikov1,r1,r2,kunz2,branikov,clement1,t1,t2,wh0,asada,potopov,abe,strong1,nandi,l1,l2,l3,l4,l5} and the wormhole stability problem \cite{zhidenko1}.  Recently, the Gibbons and Werner \cite{Gibbons}, computed the deflection angle of light applying the GBT over the optical geometry. Very soon, other consistent methods were shown to compute the deflection angles using the GBT. An interesting method for computing the deflection angle for stationary metrics was proposed by Werner \cite{werner} which involves the Finsler metric of Randers type, and recently the finite distance corrections method proposed by Ishihara et al. \cite{ishihara1,ishihara2,ishihara3}. Very recently, Crisnejo and Gallo used the GBT to study the deflection angle of massive particles by static spacetimes \cite{gabriel1,gabriel2}, while Jusufi investigated the gravitational deflection of massive particles by stationary spacetimes \cite{jusmas}.  Subsequently, this method has been applied in the context of static/rotating wormholes and topological defects \cite{K2,K22,K3,j1,j11,j2,j3,j33,j4,j5}, including the recent work by Ono et al \cite{ono}. It is worth noting that the problem the exotic matter can be avoided in some modified theories of gravity in which the wormholes are supported by matter that satisfies the energy conditions \cite{cap1,cap2,ra3,en1,en2,en3,en4,en5,en6}.

An interesting wormhole metric with toroidal topology, known as a ringhole, was proposed by Gonzalez-Diaz \cite{GonzalezDiaz:1996sr,GonzalezDiaz:1997xc}. In addition, several interesting aspects of such objects were examined; such as the possibility of observing other universe through ringholes \cite{GonzalezDiaz:2011ke,Gonzalez-Diaz:2011sia}, thermal processes in ringholes \cite{GonzalezDiaz:2010zz}, ringholes in a dark - energy universe \cite{GonzalezDiaz:2005sh,GonzalezDiaz:2003pb}. In this paper, we shall consider the gravitational lensing effect generated by the gravitational field of a ringhole, in particular we shall compute the deflection angle of light/massive particles using the GBT. 

This paper is organized as follows. In Section II, we briefly review  the ringhole metric then we shall use the GBT to determine the surface topology of rinholes. In Section III, we study the deflection of light applying GBT to the optical ringhole geometry. In Section IV, we study the deflection angle of relativistic massive particles. The last Section is devoted to conclusions.

\section{Topology of Ringholes} 
In a recent work \cite{GonzalezDiaz:1996sr,GonzalezDiaz:1997xc}, a new type of wormhole metric having the toroidal topology known as a static ringhole was introduced in terms of the following spacetime metric
\begin{equation}
ds^2=-dt^2+\left(\frac{n}{r}\right)^2 dl^2+m^2 d\varphi_1^2+(l^2+b_0^2)d\varphi_2^2,
\end{equation}
where $ -\infty <t< \infty$, $ -\infty <l< \infty$, $ 0 \leq \varphi_1, \varphi_2 \leq 2 \pi $, in which $a$ and $b$ are the radius of the circumference generated by the circular axis of the torus and that of a torus section,
respectively, while $l$ is the proper radial distance of each transversal section of the torus \cite{GonzalezDiaz:1996sr}
\begin{eqnarray}
m&=&a-(l^2+b_0^2)^{1/2}\cos \varphi_2,\\
n&=&(l^2+b_0^2)^{1/2}-a\cos \varphi_2,\\
r&=&\sqrt{ a^2+l^2+b_0^2-2 a (l^2+b_0^2)^{1/2} \cos \varphi_2},
\end{eqnarray}
with $l= \pm \sqrt{b^2-b_0^2}$, where the minus sign applies on the left side of the throat and
the plus sign does on the right side, implying $l<b$. Furthermore the maximum and the minimum circumference slices are found if we set $\varphi_2 =\pi,0$. It is possible, however, to determine the surface topology inversely. Towards this purpose, we can make use of the metric (2.1) and the GBT to show that in fact the surface topology at the wormhole throat is a torus. At a fixed moment in time $t$, and a constant $l=cosnt$, the ringhole metric reduces to 
\begin{equation}
ds^2=(a-(l^2+b_0^2)^{1/2}\cos \varphi_2)^2 d\varphi_1^2+(l^2+b_0^2)d\varphi_2^2,
\end{equation}
with the following metric tensor components
\begin{eqnarray}
g_{11}&=& (a-(l^2+b_0^2)^{1/2}\cos \varphi_2)^2,\\
g_{22}&=& l^2+b_0^2,
\end{eqnarray}
and the determinant 
\begin{eqnarray}
\det g^{(2)}&=&(a-(l^2+b_0^2)^{1/2}\cos \varphi_2)^2(l^2+b_0^2).
\end{eqnarray}

\textbf{Theorem}: \textit{Let $\mathcal{M}$ be a compact orientable surface, and let $K$ be the Gaussian curvature with respect to $g^{(2)}$ on $\mathcal{M}$. Then, the Gauss-Bonnet states that }
\begin{equation}
\iint_{\mathcal{M}} K dA=2 \pi \chi(\mathcal{M}).
\end{equation}

Note that $dA$ is the surface line element of the 2-dimensional surface and $\chi(\mathcal{M})$ is the Euler characteristic number. It is convenient to express sometimes the above theorem in terms of the Ricci scalar, in particular for the 2-dimensional surface there is a simple relation between the Gaussian curvature and Ricci scalar given by
\begin{equation}
K=\frac{\mathcal{R}}{2}.
\end{equation}

Yielding the following from 
\begin{equation}
\frac{1}{4 \pi}\iint_{\mathcal{M}} \mathcal{R} dA=\chi(\mathcal{M}).
\end{equation}

In other words, we can use the metric (2.5) and the GBT to find the Euler characteristic number. A straightforward calculation using the ringhole metric (2.5) yields the following result for the Ricci scalar
\begin{widetext}
\begin{equation}
\mathcal{R}=\frac{2 \cos\varphi_2 \left[\cos^3 \varphi_2 b_0^4 +2 \cos^2 \varphi_2 b_0^2 l^2 +\cos^3 \varphi_2 l^4 -3a \cos^2 \varphi_2 (b_0^2+l^2)^{3/2}+3 \cos \varphi_2 a^2 b_0^2 +3 \cos^3 \varphi_2 a^2 l^2 -\sqrt{b_0^2+l^2} a^3 \right]}{(b_0^2+l^2) \left(\cos^4 \varphi_2 b_0^4 +2 \cos^4 \varphi_2 b_0^2 l^2 +\cos^4 \varphi_2 l^4 -4a \cos^3 \varphi_2 (b_0^2+l^2)^{3/2}+\Delta \right)},
\end{equation}
\end{widetext}
where 
\begin{equation}
\Delta=6 \cos \varphi_2 a^2 b_0^2 +6 \cos^2 \varphi_2 a^2 l^2 -4\sqrt{b_0^2+l^2} a^3 +a^4.
\end{equation}

At the wormhole throat $l=0$, the Ricci scalar is simplified as follows 
\begin{equation}
\mathcal{R}|_{l=0}=\frac{2 \cos\varphi_2 }{b_0( b_0 \cos \varphi_2-a)}.
\end{equation}

Therefore, at the wormhole throat $l=0$, the throat radius is given by $r_{th}=b_0$; consequently from the GBT we find 
\begin{equation}
\frac{1}{4 \pi}\int_{0}^{2\pi}\int_{0}^{2 \pi} \left(\frac{2 \cos \varphi_2}{b_0( b_0 \cos \varphi_2-a)}\right)\sqrt{g^{(2)}} d \varphi_1 d \varphi_2=\chi(\mathcal{M}).
\end{equation}

Evaluating the integral we find that the Euler characteristic number is zero
\begin{equation}
\chi(\mathcal{M})=0.
\end{equation}

\begin{figure}[h!]
\center\includegraphics[width=0.43\textwidth]{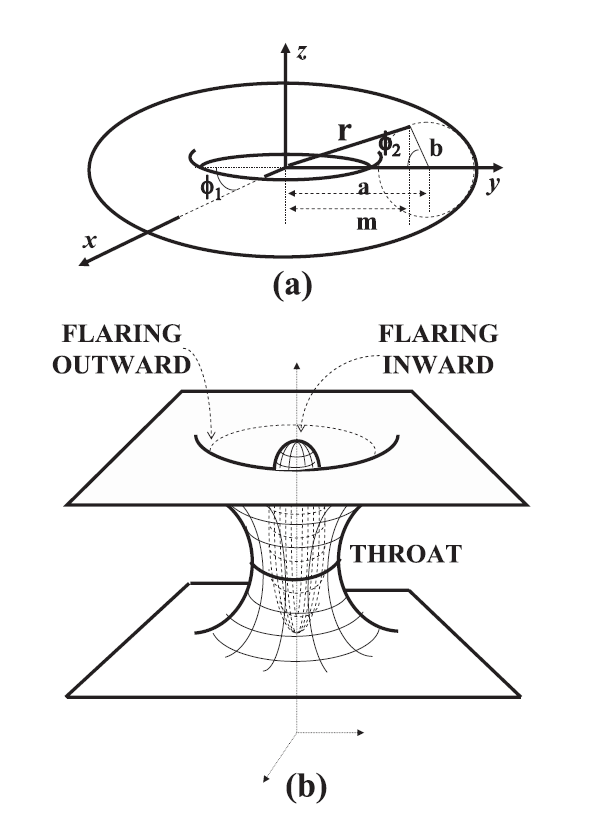}
\caption{a) Shows the toroidal structure of the ringhole throat in terms of the geometrical parameters. 
(b) Embedding diagram of a ringhole
connecting two asymptotically flat regions. Note that, the surface partly flares outward, and partly flares inward. The figure has been taken from Ref. \cite{GonzalezDiaz:2010zz}.}
\label{w3}
\end{figure}

Thus, the surface topology of the ringhole at the wormhole throat is indeed a torus, due to the simple fact that $\chi(M)_{torus}=0$. It is worth noting that metric (2.1) can be regarded as a generalization to toroidal symmetry from the static, spherical wormhole metric. Namely, performing new coordinate transformations via $a \to 0$, $\varphi_2 \to \theta+\pi/2$, $\varphi_1 \to \phi$, where  $\theta$ and $\phi$ are the angular polar coordinates on the two-sphere, one finds the Ellis wormhole solution
\begin{equation}
ds^2=-dt^2+dl^2+(l^2+b_0^2)(d\theta^2+\sin^2 \theta d\phi^2).
\end{equation}

Note that the coordinates are defined in the range $- \infty < t < \infty$, $-\infty <l < \infty $, $0 \leq \theta \leq \pi $, and $ 0 \leq \phi \leq 2 \pi $ and $b_0$ is a positive constant. Again, for a fixed moment in time $t=const$, and constant $l=const$, the above metric reduces to 
\begin{equation}
ds^2=(l^2+b_0^2)(d\theta^2+\sin^2 \theta d\phi^2),
\end{equation}
with metric tensor components 
\begin{eqnarray}
g_{11}&=&b_0^2+l^2, \\
g_{22}&=& \sin^2 \theta (l^2+b_0^2),
\end{eqnarray}
and the determinant of the metric tensor
\begin{eqnarray}
\det g^{(2)}=\sin^2 \theta (l^2+b_0^2)^2.
\end{eqnarray}

A direct calculation reveals the following result for the Ricci scalar 
\begin{equation}
\mathcal{R}=\frac{2}{l^2+b_0^2}.
\end{equation}

Again, at the wormhole throat which has the radius $r_{th}=b_0$, we find
\begin{equation}
\frac{1}{4 \pi}\int_{0}^{\pi}\int_{0}^{2 \pi} \left[\frac{2}{l^2+b_0^2} \right]_{l=0}\sqrt{g^{(2)}} d\theta d\phi=\chi(\mathcal{M}).
\end{equation}

Finally, solving the integral we find
\begin{equation}
 \chi(\mathcal{M})=2.
\end{equation}

As was expected, the surface topology of Ellis wormhole at the wormhole throat is a 2-sphere, since we know that $\chi(\mathcal{M})_{sphere}=2$. 
\vspace{0.2 cm}
\section{Deflection of light}
\subsection{Ringhole optical metric}
We can now proceed to elaborate the gravitational lensing effect in the spacetime of the ringhole metric. Let us first find the ringhole  optical metric by letting $\mathrm{d}s^2=0$, resulting with
\begin{equation}
\mathrm{d}t^{2}=\left(\frac{n}{r}\right)^2 dl^2+m^2 d\varphi_1^2+(l^2+b_0^2)d\varphi_2^2.
\end{equation}

To simplify the problem, we shall focus on the special case $\varphi_2=\pi$, which is equivalent to the problem of studying the deflection of light rays in the equatorial plane. Now the ringhole optical metric simplifies to 
\begin{eqnarray}\notag
\mathrm{d}t^{2}&=&\frac{\left((l^2+b_0^2)^{1/2}+a\right)^2}{a^2+l^2+b_0^2+2 a (l^2+b_0^2)^{1/2}} dl^2\\
&+&\left(a+(l^2+b_0^2)^{1/2}\right)^2d\varphi_1^2.
\end{eqnarray}

We can easily read the optical metric components from the last equation as follows
\begin{eqnarray}
g_{11}^{(op)}&=& \frac{\left((l^2+b_0^2)^{1/2}+a\right)^2}{a^2+l^2+b_0^2+2 a (l^2+b_0^2)^{1/2}},  \\
g_{22}^{(op)}&=& \left(a+(l^2+b_0^2)^{1/2}\right)^2.
\end{eqnarray}

And the determinant of the ringhole optical metric is computed as follows
\begin{equation}
\det g^{(op)}=a^2+l^2+b_0^2+2 a \sqrt{b_0^2+l^2}.
\end{equation}

The Gaussian optical curvature $K$ can be found from the optical metric (3.2). To do this, first we can find the Ricci scalar $\mathcal{R}^{(op)}$ from the optical metric (3.2), then using the relation $K=\mathcal{R}^{(op)}/2$, we find
\begin{widetext}
\begin{eqnarray}
K&=& - \frac{\left(b_0^2+l^2\right)^{3/2} \left[3 a^2 b_0^2+4 a^2 l^2 +b_0^4+2b_0^2 l^2+l^4\right]-\sqrt{b_0^2+l^2} \left[4a^2 b_0^2l^2+a^2 l^4+b_0^4 l^4 +2 b_0^2 l^4+l^6 \right]+\eta  }{\left(a^2+l^2+b_0^2+2 a \sqrt{b_0^2+l^2}\right)^2\left(b_0^2+l^2\right)^{5/2}},
\end{eqnarray}
\end{widetext}
where 
\begin{equation}
\eta=a^3 b_0^4+a^3b_0^2 l^2+3 a b_0^6+6a b_0^4 l^2+3 a b_0^2 l^4.
\end{equation}

Introducing dimensionless variables, say, $x=a/l$ and $y=b_0/l$, then performing a series expansion in terms of $x$ and $y$, the Gaussian optical curvature can be approximated in leading order terms as follows
\begin{equation}
K\simeq - \left(\frac{l-a}{l^5}\right)b_0^2+\mathcal{O}\left(\frac{b_0^2 a^2}{l^6}\right).
\end{equation}

\subsection{Deflection angle of light}

\textbf{Theorem}: \textit{Let $\mathcal{A}_{R}$  be a non-singular domain (or a region outside the light ray) with boundaries $\partial 
\mathcal{A}_{R}=\gamma_{g^{(op)}}\cup C_{R}$, of an oriented two-dimensional surface $S$ with the optical metric $g^{(op)}$. Let $K$ and $\kappa $ be the Gaussian optical
curvature and the geodesic curvature, respectively. Then, the GBT in terms of the above construction is written as follows \cite{Gibbons}}
\begin{equation}
\iint\limits_{\mathcal{A}_{R}}K\,\mathrm{d}S+\oint\limits_{\partial \mathcal{%
A}_{R}}\kappa \,\mathrm{d}t+\sum_{k}\theta _{k}=2\pi \chi (\mathcal{A}_{R}).
\label{10}
\end{equation}

Where $\mathrm{d}S$ is the optical surface element, $\theta _{k}$ gives the exterior angle at the $k^{th}$ vertex. The regular domain is chosen to be outside of the light
ray in the $(r,\varphi)$ optical plane, this domain can be thought to have the topology of disc with the  Euler characteristic number $\chi (\mathcal{A}_{R})=1$. Let us introduce a smooth curve defined as $\gamma:=\{t\}\to \mathcal{A}_{R}$,  with the geodesic curvature defined by the following relation
\begin{equation}
\kappa =g^{(op)}\,\left( \nabla _{\dot{\gamma}}\dot{\gamma},\ddot{\gamma}%
\right),  
\end{equation}%
 with an additional unit speed condition $g^{(op)}(\dot{\gamma},\dot{\gamma})=1$, and $\ddot{\gamma}$ being the unit acceleration vector. Now if we consider a very large, but finite radial distance $l\equiv R\rightarrow \infty $, such that the two jump angles (at the source $\mathcal{S}$, and observer $\mathcal{O})
$, yields  $\theta _{\mathit{O}}+\theta _{\mathit{S}}\rightarrow \pi $ \cite{Gibbons}. Note that, by definition, the geodesic curvature for the light ray (geodesics) $\gamma_{g^{(op)}}$ vanishes, i.e. $\kappa (\gamma_{g^{(op)}})=0$. One should only compute the contribution to the curve $C_{R}$. That being said, from the GBT we find
\begin{equation}
\lim_{R\rightarrow \infty }\int_{0}^{\pi+\hat{\alpha}}\left[\kappa \frac{d t}{d \varphi_1}\right]_{C_R} d \varphi_1=\pi-\lim_{R\rightarrow \infty }\iint\limits_{\mathcal{A}_{R}}K\,\mathrm{d}S
\end{equation}

The geodesic curvature for the curve $C_{R}$ located at a coordinate distance $R$ from the coordinate system chosen at the ringhole center can be calculated in terms of the relation
\begin{equation}
\kappa (C_{R})=|\nabla _{\dot{C}_{R}}\dot{C}_{R}|.
\end{equation}

In components notation we can write 
\begin{equation}
\left( \nabla _{\dot{C}_{R}}\dot{C}_{R}\right) ^{r}=\dot{C}_{R}^{\varphi_1
}\,\left( \partial _{\varphi_1 }\dot{C}_{R}^{r}\right) +\Gamma%
_{\varphi_1 \varphi_1 }^{r(op)}\left( \dot{C}_{R}^{\varphi_1 }\right) ^{2}. 
\end{equation}

With the help of the unit speed condition and the ringhole optical metric one can show that
\begin{eqnarray}
\lim_{R\rightarrow \infty }\kappa (C_{R}) &=&\lim_{R\rightarrow \infty
}\left\vert \nabla _{\dot{C}_{R}}\dot{C}_{R}\right\vert ,  \notag \\
&\rightarrow &\frac{1}{R}.
\end{eqnarray}

On the other hand, from the ringhole optical metric setting a constant $l\equiv R$, we find 
\begin{eqnarray}
\lim_{R\rightarrow \infty }\mathrm{d}t \rightarrow R\,\mathrm{d}\varphi_1.
\end{eqnarray}

Combining these two equations, we arrive at the conclusion that our ringhole optical metric is asymptotically Euclidean 
\begin{eqnarray}
\lim_{R\rightarrow \infty } \left(\kappa (C_{R})\frac{\mathrm{d}t}{\mathrm{d}\varphi_1}\right)=1.
\end{eqnarray}

From the GBT (3.11) it is not difficult to solve for the deflection angle which gives
\begin{equation}
\hat{\alpha}=-\int\limits_{0}^{\pi }\int\limits_{r(\varphi_1)%
}^{\infty } K \mathrm{d}S. 
\end{equation}
where the surface element reads $\mathrm{d}S =\sqrt{g^{(op)}}dl d\varphi_1$. Moreover we shall assume the following equation for the light ray $r(\varphi_1)=\mathsf{b}/\sin \varphi_1 $. Using expression (3.8) we find 
\begin{equation}
\hat{\alpha}=-\int\limits_{0}^{\pi }\int\limits_{\frac{\mathsf{b}}{\sin \varphi_1 }%
}^{\infty }\left[- \left(\frac{l-a}{l^5}\right)b_0^2 \right]\mathrm{d}S. 
\end{equation}
in which the surface element can be approximated as
\begin{equation}
\mathrm{d}S  \simeq (l+a) dl d\varphi_1.
\end{equation}

Solving this integral we can approximate the solution to be
\begin{equation}
\hat{\alpha}= \frac{ \pi b_0^2}{4\, \mathsf{b}^2}-\frac{ 3 b_0^2 a^2 \pi}{32 \mathsf{b}^4}.
\end{equation}

As a special case, for a vanishing $a$, i.e., $a \to 0$, we recover the deflection angle by Ellis wormhole 
\begin{equation}
\hat{\alpha}_{Ellis}= \frac{ \pi b_0^2}{4\, \mathsf{b}^2}.
\end{equation}

\section{Deflection of massive particles}

In this section we shall consider the problem of computing the deflection angle for relativistic massive particles. For the same reason,  let us consider the physical spacetime metric to be described by a the following metric \cite{gabriel1}
\begin{equation}
ds^2=-A(r)dt^2+B(r)dr^2+C(r)(d\theta^2+\sin^2 \theta d\varphi^2).
\end{equation}

We assume that the particles has speed $v$ and energy  \cite{gabriel1}
\begin{equation}
E_{\infty}=\frac{\mu}{\sqrt{1-v^2}},
\end{equation}
as measured by an asymptotic observer. In addition, let us assume that the particle has an angular momentum given by
\begin{equation}
J=\frac{\mu v \mathsf{b}}{\sqrt{1-v^2}},
\end{equation}
where $\mathsf{b}$ is the impact parameter. Without going into details (see for details \cite{gabriel1}), one can deduce the following optical metric for massive particles 
\begin{equation}
d\sigma^2=\frac{n(r)^2}{A(r)}\left(B(r)dr^2+C(r) d\varphi^2 \right),
\end{equation}
where the refractive index is given by
\begin{equation}
n^2(r)=1-\frac{\mu^2}{E^2_{\infty}}A(r)=1-(1-v^2)A(r).
\end{equation}

In our particular case, we can recast the ringhole metric in the form (4.4) by choosing $\varphi_2=\pi$, yielding 
\begin{eqnarray}\notag
\mathrm{d}\sigma^{2}&=&\left((1-(1-v^2)\right)\Big[\frac{\left((l^2+b_0^2)^{1/2}+a\right)^2}{a^2+l^2+b_0^2+2 a (l^2+b_0^2)^{1/2}} dl^2 \\
&+&\left(a+(l^2+b_0^2)^{1/2}\right)^2d\varphi_1^2\Big].
\end{eqnarray}
note that in our case $A=1$. The Gaussian optical curvature from the last metric is computed as follows
\begin{widetext}
\begin{eqnarray}
K&=& - \frac{\left(b_0^2+l^2\right)^{3/2} \left[3 a^2 b_0^2+4 a^2 l^2 +b_0^4+2b_0^2 l^2+l^4\right]-\sqrt{b_0^2+l^2} \left[4a^2 b_0^2l^2+a^2 l^4+b_0^4 l^4 +2 b_0^2 l^4+l^6 \right]+\eta  }{v^2\left(a^2+l^2+b_0^2+2 a \sqrt{b_0^2+l^2}\right)^2\left(b_0^2+l^2\right)^{5/2}}.
\end{eqnarray}
\end{widetext}

Or, if we perform a series expansion around $b_0$, $a$, we get
\begin{equation}
K\simeq - \left(\frac{l-a}{v^2 l^5}\right)b_0^2.
\end{equation}

It is clearly seen from the last equation that, in the limit $v \to 0$, there is a singularity in $K$. Due to this singularity, we need to add a further constraint, namely the speed of the particle belongs to the interval $0<v \leq 1$ with [$c=1$].

Next, we find that the geodesic deviation for large coordinate radius $R$ reads
\begin{eqnarray}
\lim_{R\rightarrow \infty }\kappa (C_{R}) &=&\lim_{R\rightarrow \infty
}\left\vert \nabla _{\dot{C}_{R}}\dot{C}_{R}\right\vert ,  \notag \\
&\rightarrow &\frac{1}{v \,R}.
\end{eqnarray}

From the metric (4.6) in this limit for a constant $l=R$ we also find that
\begin{eqnarray}
\lim_{R\rightarrow \infty }\mathrm{d}\sigma \rightarrow v \,R\,\mathrm{d}\varphi_1.
\end{eqnarray}

From these equations it is possible to see that same condition (3.16) holds, namely
\begin{eqnarray}
\lim_{R\rightarrow \infty }\left(\kappa(C_R)\frac{\mathrm{d}\sigma}{\mathrm{d}\varphi_1}\right)=1.
\end{eqnarray}

Therefore the deflection angle is found to be
\begin{equation}
\hat{\alpha}=-\int\limits_{0}^{\pi }\int\limits_{\frac{\mathsf{b}}{\sin \varphi }%
}^{\infty }\left[- \left(\frac{l-a}{v^2 l^5}\right)b_0^2 \right]\mathrm{d}S. 
\end{equation}
where the surface element can be approximated as
\begin{equation}
\mathrm{d}S \simeq v^2 (l+a) dl d\varphi_1.
\end{equation}

Finally, solving this integral we find
\begin{equation}
\hat{\alpha} \simeq \frac{ \pi b_0^2}{4\, \mathsf{b}^2}-\frac{ 3 b_0^2 a^2 \pi}{32 \mathsf{b}^4}.
\end{equation}

In other words, we recovered the same result as in the case of deflection of light.   However, this is not a surprising result considering the fact that the deflection angle depends only on the geometric parameters encoded via $b_0$ and $a$ reflecting the toroidal structure of the ringhole. 

\section{Conclusions}

Let us summarize the main results of this brief review. We have studied the surface topology and the deflection angle of light in the ringhole spacetime using GBT. Using the ringhole metric, we have shown that the surface topology of the ringhole at the throat is indeed a torus with the Euler characteristic number $\chi_{torus}=0$. The peculiar ringhole geometry under specific coordinate transformations reduces to the Ellis wormhole which has the surface topology of a 2-sphere, with Euler characteristic number $\chi_{sphere}=2$. 

Most importantly, we carried out a detailed study of the deflection angle of light caused by the ringhole geometry in the weak limit approximation. Among other things, we have shown that the deflection angle is given in terms of the ringhole throat radius $b_0$, and the radius of the circumference generated by the circular axis of the torus $a$. In particular we found the following equation for the deflection angle
\begin{equation}\notag
\hat{\alpha}_{Ringholes} \simeq \frac{ \pi b_0^2}{4\, \mathsf{b}^2}-\frac{ 3 b_0^2 a^2 \pi}{32 \mathsf{b}^4}.
\end{equation}

We observe that the contribution of the fist term (which is proportional to the wormhole throat $b_0$), indicating  that light rays always bend towards the ringhole, while the effect of the second term  (which is proportional to the radius $b_0^2 a^2$), deflects the light rays outwards the ringhole. This result can be viewed as a generalization of the deflection angle caused by Ellis wormholes simply by setting $a=0$.  A further analyses shows that the same result is deduced  by studying the deflection angle of massive particles moving with a relativistic velocity $v$.  

Finally, we should point out that the resulting difference between the ringhole deflection angle with the Ellis wormholes can be of significant importance in astrophysics for the following reason: we can potentially use this difference in the deflection angle to distinguish ringholes from wormholes and black holes. Note that, there is another effect pointed out in \cite{GonzalezDiaz:2011ke,Gonzalez-Diaz:2011sia}  where two concentric bright rings can be produced by the inner gravitational field of a ringhole with toroidal symmetry when a single luminous source is situated behind the ringhole in our universe or in a parallel universe. This might be a signature of the ringhole spacetime and can be observed by the Earth observer. This effect can also  distinguish them from black holes or other exotic objects.  We are planning in the near future to investigate the deflection angle of massive particles by adding the effect of rotation in the ringhole geometry.


\begin{thebibliography}{999}

\bibitem{Flamm} L. Flamm,  Phys. Z. 17, 448 (1916).

\bibitem{Einstein} A. Einstein and N. Rosen, Phys. Rev.  \textbf{48},
73 (1935).

\bibitem{fw} R. W. Fuller and J. A. Wheeler, Phys. Rev. \textbf{128}, 919 (1962).

\bibitem{wh1} J. Wheeler, Ann. Phys. \textbf{2}, (6) 604 (1957).

\bibitem{Wh2} J. A. Wheeler, Phys. Rev. \textbf{97}, 511 (1955).

\bibitem{Wh3} J. A. Wheeler, \textit{Geometrodynamics}, Academic Press, New
York, 1962.

\bibitem{Ellis1} H. G. Ellis, J. Math. Phys. \textbf{14}, 104 (1973).

\bibitem{Ellis2} H. G. Ellis, J. Math. Phys. \textbf{15}, 520 (1974)(Erratum).

\bibitem{br1} K.A. Bronnikov, Acta Phys.Polon. B4 (1973) 251-266

\bibitem{clm}  G.~Clement,  Gen. Rel. Grav. \textbf{16}, 131 (1984)

\bibitem{Morris} M. S. Morris and K. S. Thorne, Am. J. Phys. \textbf{56}, 395 (1988).

\bibitem{Visser} M. Visser, \textit{Lorentzian Wormholes} (AIP Press, New York,
1996).


\bibitem{Teo} E. Teo, Phys. Rev. D \textbf{58}, 024014 (1998).

\bibitem{lobo1} Francisco S. N. Lobo, Phys.Rev.D71:084011,2005

\bibitem{lobo2} Jose' P. S. Lemos, Francisco S. N. Lobo, Sergio Quinet de Oliveira, Phys.Rev.D68:064004,2003

\bibitem{branikov1} K.A. Bronnikov, A.M. Galiakhmetov,  Phys. Rev. D 94, 124006 (2016)

 \bibitem{r1}  R. Shaikh,  Phys. Rev. D 92, 024015 (2015)
  
  \bibitem{r2} R. Shaikh, Sayan Kar, Phys. Rev. D 96, 044037 (2017)
  
\bibitem{kunz2} V. Dzhunushaliev, V. Folomeev, B. Kleihaus, J. Kunz, Phys. Rev. D 97, 024002 (2018)

\bibitem{branikov} K.A. Bronnikov, S.V. Chervon, S.V. Sushkov, Grav.Cosmol.15:241-246,2009


\bibitem{clement1}  G.~Clement,
  Phys.\ Rev.\ D {\bf 51}, 6803 (1995).
  
 
\bibitem{t1} N. Tsukamoto and T. Harada, Phys. Rev. D \textbf{95}, 024030 (2017).

\bibitem{t2} N. Tsukamoto, T. Harada and K. Yajima, Phys. Rev. D \textbf{86}, 104062 (2012).


\bibitem{wh0} L. Chetouani and G. Clement, Gen. Rel. Grav. \textbf{16}, 111-119 (1984).  

\bibitem{asada} K. Nakajima, H. Asada and Phys.Rev.D \textbf{85}, 107501 (2012).

\bibitem{potopov} A. Bhattachary, A. Potapov, Mod. Phys. Lett. A \textbf{25}, 2399 (2010). 

\bibitem{abe} F. Abe, ApJ \textbf{725} (2010) 787-793


\bibitem{strong1} T. K. Dey and S. Sen, Mod. Phys. Lett. A \textbf{23}, 953-962, (2008).


\bibitem{nandi} K. K. Nandi, Y. Zhang and A. V. Zakharov, Phys.Rev. D \textbf{74}, 024020 (2006).
  
  \bibitem{l1} M. Amir, K. Jusufi, A. Banerjee, and S. Hansraj, Class. Quantum Grav. 36, 215007 (2019)
  
  \bibitem{l2} A. Abdujabbarov, B. Juraev, B. Ahmedov, and Z. Stuchlik, Astrophys. Space Sci. 361, (2016) 226.
  
  \bibitem{l3} C. Bambi, Phys. Rev. D 87, 084039 (2013).
  
  \bibitem{l4} R. Shaikh, Phys. Rev. D 98, 024044 (2018)

  
  \bibitem{l5} G. Gyulchev, P. Nedkova, V. Tinchev, S. Yazadjiev, Eur. Phys. J. C (2018) 78: 544
  
   \bibitem{zhidenko1} K. A. Bronnikov, R. A. Konoplya, A. Zhidenko, Phys. Rev. D 86, 024028 (2012)
   
  \bibitem{Gibbons}  G.~W.~Gibbons and M.~C.~Werner,
  Class.\ Quant.\ Grav.\  {\bf 25}, 235009 (2008).
  
  \bibitem{werner}M.~C.~Werner, Gen. Rel. Grav. 44, 3047-3057 (2012)
  
  \bibitem{ishihara1} A. Ishihara, Y. Suzuki, T. Ono, T. Kitamura, H. Asada,  Phys. Rev. D 94, 084015 (2016)
  
  \bibitem{ishihara2} A. Ishihara, Y. Suzuki, T. Ono, H. Asada,  Phys. Rev. D 95, 044017 (2017)
  
  \bibitem{ishihara3} T. Ono, A. Ishihara, H. Asada,   Phys. Rev. D 96, 104037 (2017)
 
 \bibitem{gabriel1} G. Crisnejo, E. Gallo, Phys. Rev. D 97, 124016 (2018)
 
  \bibitem{gabriel2}  G. Crisnejo, E. Gallo, A. Rogers, Phys. Rev. D 99, 124001 (2019)

 
 \bibitem{jusmas}	K. Jusufi, Phys. Rev. D 98, 064017 (2018)
  \bibitem{K2} 
  K.~Jusufi,
  Int.\ J.\ Geom.\ Meth.\ Mod.\ Phys.\  {\bf 14}, no. 12, 1750179 (2017).

 \bibitem{K22}  K. Jusufi, Phys. Rev. D 98, 044016 (2018)

   \bibitem{K3} 
  K.~Jusufi and A.~\"{O}vg\"{u}n,
  Phys.\ Rev.\ D {\bf 97}, no. 2, 024042 (2018).
 
\bibitem{j1} 
  K.~Jusufi,
  Astrophys.\ Space Sci.\  {\bf 361}, no. 1, 24 (2016)
  \bibitem{j11} 
  K.~Jusufi, F.~Rahaman and A.~Banerjee,
  Annals Phys.\  {\bf 389}, 219 (2018)
  
\bibitem{j2} 
  K.~Jusufi and A.~\"{O}vg\"{u}n,
  Phys.\ Rev.\ D {\bf 97}, no. 6, 064030 (2018)
  
\bibitem{j3} 
  K.~Jusufi, A.~\"{O}vg\"{u}n, and A.~Banerjee,
  Phys.\ Rev.\ D {\bf 96}, no. 8, 084036 (2017)
  \bibitem{j33}  K.~Jusufi, N.~Sarkar, F.~Rahaman, A.~Banerjee and S.~Hansraj,
  Eur. Phys. J. C (2018) 78: 349.
  
\bibitem{j4}
  K.~Jusufi, I.~Sakalli and A.~Ovgun,
  Phys.\ Rev.\ D {\bf 96} (2017) no.2,  024040
  
\bibitem{j5} 
  K.~Jusufi, M.~C.~Werner, A.~Banerjee and A.~Ovgun,
  Phys.\ Rev.\ D {\bf 95}, no. 10, 104012 (2017)
  
  \bibitem{ono} T. Ono, A. Ishihara, H.i Asada, Phys. Rev. D 98, 044047 (2018)

\bibitem{cap1} S. Capozziello, F. S. N. Lobo, and J. P. Mimoso, Phys. Rev D 91, 124019 (2015).

\bibitem{cap2} S. Capozziello, F. S. N. Lobo, and J. P. Mimoso, Phys. Lett. B 730, 280 (2014).

\bibitem{ra3} R. Shaikh, Phys. Rev. D 98, 064033 (2018)

\bibitem{en1} H. Maeda and M. Nozawa, Phys. Rev. D 78, 024005 (2008).

\bibitem{en2} F. S. N. Lobo and M. A. Oliveira, Phys. Rev. D 80, 104012 (2009).

\bibitem{en3} M. H. Dehghani and S. H. Hendi, Gen. Relativ. Gravit. 41, 1853 (2009).

\bibitem{en4} P. Kanti, B. Kleihaus, and J. Kunz, Phys. Rev. Lett. 107, 271101 (2011).

\bibitem{en5}N. M. Garcia and F. S. N. Lobo, Class. Quant. Grav. 28, 085018 (2011).

\bibitem{en6} C. G. Boehmer, T. Harko, and F. S. N. Lobo, Phys. Rev. D 85, 044033 (2012).



\bibitem{GonzalezDiaz:1996sr}
  P.~F.~Gonzalez-Diaz,
  Phys.\ Rev.\ D {\bf 54} (1996) 6122
  [gr-qc/9608059].

\bibitem{GonzalezDiaz:1997xc}
  P.~F.~Gonzalez-Diaz,
  Phys.\ Rev.\ D {\bf 56} (1997) 6293
  [gr-qc/9708044].
  
\bibitem{GonzalezDiaz:2011ke}
  P.~F.~Gonzalez-Diaz and A.~Alonso-Serrano,
  Phys.\ Rev.\ D {\bf 84} (2011) 023008
  [arXiv:1102.3784 [astro-ph.CO]].
\bibitem{Gonzalez-Diaz:2011sia}
  P.~F.~González-Díaz,
  Springer Proc.\ Phys.\  {\bf 137} (2011) 193.
\bibitem{GonzalezDiaz:2010zz}
  P.~F.~Gonzalez-Diaz,
  Phys.\ Rev.\ D {\bf 82} (2010) 044016.
\bibitem{GonzalezDiaz:2005sh}
  P.~F.~Gonzalez-Diaz,
  Grav.\ Cosmol.\  {\bf 12} (2006) 29
  [astro-ph/0507714].
\bibitem{GonzalezDiaz:2003pb}
  P.~F.~Gonzalez-Diaz,
  Phys.\ Rev.\ D {\bf 68} (2003) 084016
  [astro-ph/0308382].
 
  

\end{thebibliography}
\end{document}